# APPLICATION OF DATA MINING IN PROTEIN SEQUENCE CLASSIFICATION

Suprativ Saha[1] and Rituparna Chaki[2]

[1]Department of Computer Science & Engineering, Global Institute of Management and Technology, Krishnagar City, West Bengal, India
reach2suprativ@yahoo.co.in
[2]Department of Computer Science & Engineering, West Bengal University of Technology, Saltlake, Kolkata, India
rituchaki@gmail.com

## ABSTRACT

*Protein sequence classification involves feature selection for accurate classification. Popular protein sequence classification techniques involve extraction of specific features from the sequences. Researchers apply some well-known classification techniques like neural networks, Genetic algorithm, Fuzzy ARTMAP, Rough Set Classifier etc for accurate classification. This paper presents a review is with three different classification models such as neural network model, fuzzy ARTMAP model and Rough set classifier model. This is followed by a new technique for classifying protein sequences. The proposed model is typically implemented with an own designed tool and tries to reduce the computational overheads encountered by earlier approaches and increase the accuracy of classification.*

## 1. INTRODUCTION

The introduction of new technologies such as computers, satellites and many others has lead to an exponential growth of collected data in many areas. Traditional data analysis techniques often fail to process large amounts of data efficiently. In this case data mining technology can be used to extract knowledge from large amount of data. Recently, the collection of biological data like protein sequences, DNA sequences etc. is increasing at explosive rate due to improvements of existing technologies and the introduction of new ones such as the microarrays. So Data mining technique is used to extract meaningful information from the huge amount of biological data sequences, such as the DNA, protein etc. One important area of research is to classify protein sequences into different families, classes or sub classes.

Classification is the most important technique to identify a particular character or a group of them. Different classification methods or algorithms have been proposed by different researchers to classify the protein sequences. The Protein sequence consists of twenty different amino acids which are arranged in some specific sequences. Popular protein sequence classification techniques involve extraction of specific features from the sequences. These features depend on the structural and functional properties of amino acids. These features are compared with their predefined values. Using neural networks, Genetic algorithm, Fuzzy ARTMAP, Rough Set based Classifier etc till date; none of researchers have achieved 100% accuracy level. This paper

International Journal of Database Management Systems ( IJDMS ) Vol.4, No.5, October 2012presents a comprehensive study of the on-going research on protein sequence classification followed by a comparative analysis.

The rest of the paper is organized as follows: Section 2 presents a review of classification models; section 3 consists of a comparative analysis, followed by a proposed work in section 4 and implementation in Section 5. Finally Section 6 presents the result discursion and Section 7 presents the conclusion.

## 2. REVIEW

Different classification techniques have been used to classify protein sequence into its particular class, sub class or family. All these methods aim to extract some features, match the value of these features and finally classify the protein sequence. This paper focuses on mainly three types of classification techniques, the (i) Neural network Model, (ii) the Fuzzy ARTMAP Model, and (iii) the Rough Set Classifier.

### 2.1. Neural Network Model

Generally there are different types of approaches available for classification, such as decision trees and neural networks. Extracted features of protein sequences are distributed in a high dimensional space with complex characteristics, which is difficult to satisfy model using some parameterized approaches. So neural network based classifier have been chosen to classify protein sequence. Decision tree based techniques fails to classify patterns with continuous features especially as the number of attributes is larger.

Neural network model [2] has been used to classify unknown protein sequences by extracting some features from it which can apply as input of this model. 2-gram encoding method and 6-letter exchange group methods were used to find global similarity. For local similarity, user defined variables Len, Mut, and occur were used. Minimum description length (MDL) principle was also used to calculate the significance of motif. Some predefined values of these features were used as intermediate layers or hidden layers of the neural network. This model produces 90% to 92% accuracy.

In [1] authors want to classify the protein sequences using neural network model. Here n-gram encoding method (n = 2, 3… N and N = len. of the input sequence) was used to extract feature which was applied to construct the pattern matrix. At the end by using neural network model new pattern was matched with the predefine pattern of protein super families or families. N-gram encoding method includes all 2-gram, 3-gram, etc encoding method, so to form the pattern matrix of features extracted from n-gram encoding method, individual also needed. Therefore in case of large sequences computational overhead also be increase. The accuracy level remains 90% only.

[7] Proposes an advance technique of [1]. At first 2-gram encoding method is applied and using only this result pattern matrix is build. If this matrix is unable to classify the input protein sequence, result of 3-gram encoding method is added to the pattern matrix. The result is then matched using neural network. The performance of this technique is largely dependent on the number of encoding operations performed. In case all the sub patterns are to be checked, performance deteriorates sharply. The average performance is slightly better than [1].

In [9] authors used a probabilistic neural network model. The paper uses self organized map (SOM) network. The SOM networks can be used to discover relationships within a set of protein sequences by clustering them into different groups. Different types of features like Amino acid distribution, 2-gram amino acid distribution, etc, were extracted from the input protein sequence to construct the pattern matrix. According to the unsupervised learning method of neural network input sequences are placed in the $1^{st}$ layer of neural network, then pattern matrix is presented in the hidden layer ($2^{nd}$ layer) for matching with some predefine values. Different outcome results are summarized in the $3^{rd}$ layer. $4^{th}$ or final layer of the probabilistic neural network model

104

produced the final result of classification. The technique failed to produce impressive results in case of $unclassified_p$ and $unclassified_n$ parameters. The use of SOM network also causes hindrance in interpreting the results.

The main limitations of SOM networks for protein sequence classification are its interpretability of the results, and the model selection. SOM is a straight forward method; there is no chance of back propagation. But to reach a particular goal and increase the accuracy level of the classification back propagation is most important technique. In back propagation based model, there is a chance to move to the previous steps.

The problems faced by the SOM based technique in [9] is overcome by back propagation neural network (BPNN) technique in [4]. Here authors use extreme learning machine to classify protein sequence. This extreme learning machine included the advancement of back propagation technique of neural network model. To evaluate the performance of this machine authors extracted some features like 2-gram encoding method and 6-letter exchange method from the input protein sequence. A pattern matrix was formed using those features and used in the extreme learning machine. Finally accuracy level also is measured.

The use of neural network technique normally neural network is good at handling non-linear data (noise data). The protein sequence being linear, use of neural network does not add up. It has been observed that sequences of 20 different amino acids (Protein sequences) were used as working data in this paper. The data being linear, the use of neural network modelling fails to add any extra benefit. The paper fails to take care of noise in protein sequence even through it uses neural network. The model failed to process the physical relationships which are most important in this purpose. Again regarding the accuracy issue, neural network model provide 90%-92% accuracy. Improvement of this accuracy is mostly needed.

## 2.2. Fuzzy ARTMAP Model

Generally Fuzzy ARTMAP model, a machine learning method is used to classify the protein sequence. The basic difference between neural network model and fuzzy model is that neural network model do not analysis the data individually, it only provide a knowledge based information. On the other hand Fuzzy model calculates the membership value of every feature using membership functions and implements it in the whole model.

This model [5] was implemented to classify the unknown protein sequence into different predefine protein families or protein sub families with 93% high accuracy. A cleaning process was also been conducted on the databases. After that different features were extracted from protein sequences, e.g. physic-chemical properties of the sequences. They calculated the molecular weight (W) and the isoelectric point (pI) of the protein sequences, followed by Amino acid composition of the sequences. The hydropathy composition (C), the hydropathy distribution (D) and the Hydropathy transmission (T) also calculated. After extracting all forty different features an unknown protein sequence was used as the input of the Fuzzy ARTMAP model. Some predefine features which were extracted or generated from known protein sequence also used as the unit of classification rules. This model generated the name of family or sub family of the unknown protein sequence as the output which was taken as the input of the Fuzzy ARTMAP model.

In [6] author wants to classify protein sequence using Fuzzy model. Calculating the membership value using the membership function is most important in fuzzy model. At first feature is extracted using 6-letter exchange group method. Then membership value is assigned and constructs the pattern matrix. Using a fuzzy rule pattern matrix was distributed into 3 small groups (i) small, (ii) medium and (iii) large. Now according to the target, choose a group and further distribution was done to reach to goal. At the end, the model is tested using uniport 11.0 dataset which contain globin, kinase, ras and trypsin super families of protein. In this paper number of antecedent variables is huge. It is right that increase of antecedent variable, increase the classification accuracy but it also increases the CPU time.

[8] Proposes an advancement of the techniques proposed in [6]. This technique tries to decrease the CPU time without changing the classification accuracy. Here features also extracted using 2-gram encoding method and 6-letter exchange group method and according to the membership value of features pattern matrix was distributed into 3 small groups. But executing the distribution method a new algorithm is applied on the value of features. This algorithm provides a rank on the value of the features using the feature ranking algorithm and according to the rank features is arranged in descending order. Now collect the top ranked features to construct the pattern matrix. In this way this technique can reduce the CPU overhead. This is a normal, easy, human understandable and alignment – independent method. As a result every biologist can easily understand this method and feel free to implement it. At the end this method is evaluated and compared to the non fuzzy technique (C 4.5). The computational complexity is reduced, but the accuracy level remains the same as the earlier method [6].

Fuzzy modeling helps in the data analysis although storage and time requirement are high. The construction of fuzzy sets, for every iteration adds up to the computational complexity as well. This model also failed to process the physical relationships which are most important in this purpose.

## 2.3. Rough Set Classifier

Generally machine learning methods such as the neural network model, Fuzzy ARTMAP model etc., are insufficient to handle large number of unnecessary features, extracted for rule discovery [11]. As a result they try to select the features to reduce the computation time. But these methods also degrade their performance. Accuracy level is not sufficient since every feature is equally important for proper classification. Rough set classifier is a new model to overcome this problem.

Rough sets theory is a machine learning method, which is introduced by Pawlak [3] in the early 1980s. It implements the concept of set theory to make some decision. The indiscernibility relation that induced minimal decision rules from training examples is the important notation in rough set model. To identify the minimal set of the features, if-else rule is used on the decision table.

This new classification model [10] can classify the voluminous protein data based on structural and functional properties of protein. This model is faster, accurate and efficient classification tool than the others. Rough Set Protein Classifier provides 97.7% accuracy. It is a hybridized tool comprising Sequence Arithmetic, Rough Set Theory and Concept Lattice. It reduces the domain search space to 9% without losing the potentiality of classification of proteins. An innovative technique viz., Sequence Arithmetic (SA) to identify family information and utilize it for reducing the domain search space is proposed. Rules are generated and stored in Sequence Arithmetic database. A new approach to compute predominant attributes (approximate reducts) and use them to construct decision tree called Reduce based Decision Tree (RDT) is proposed. Decision rules generated from the RDT are stored in RDT Rules Database (RDTRD). These rules are used to obtain class information. The infirmity of RDT is overcome by extracting spatial information by means of Neighbourhood Analysis (NA). Spatial information is converted into binary information using threshold. It is utilized for the construction of Concept Lattice (CL). The Associated Rules from the CL are stored in Concept lattice Association Rule Database (CARD). Further, the domain search space is confined to a set of sequences within a class by using these Association Rules. Time complexity of this model is $O(n) + O(f) + O(\log C) + O(2^r)$, where the unknown sequence y with size n. No of the families in the database is f. 'C' is the number of classes in a given family and 'r' is the number of proteins in classes then the CL will have $2^r$ nodes.

In [11] authors use rough set classifier to extract all the features necessary for classification. The feature set was built based on compositional percentages of the 20 amino acids properties. The authors had used Rosetta system for data mining and knowledge discovery. In the first phase, a method is implemented on the whole datasets in which all the subfamilies were included ignoring the small size of sequences. Rough set model generally use standard Genetic Algorithms. The Rough Set was further applied to classify the data and evaluate the performance of the seven subfamilies. This paper achieves a satisfactory accuracy level without increase the computational time.

The Rough Set Classifier model provides knowledge based information only without any analysis of data. For properly classifying protein sequences, both play an important role. Instead of classifying protein sequence into classes or sub classes, this model provides a small known sequence from a long unknown protein sequence. Thus it requires extra time and space for further classification of the output sequence into classes or sub classes. The accuracy level is 97.7%, which leaves scope for improvement.

## 3. COMPARATIVE ANALYSIS

| Techniques | Neural Network based Classifier [1,2,4,7,9] | Fuzzy ARTMAP based Classifier [5,6,8] | Rough Set based Classifier [3,10,11] |
|---|---|---|---|
| Database Uses | The Int. Protein Seq. Database Release 62 | i) SCOP 1.69<br>ii) ASTRAL 1.69 | NCBI (Blast) |
| Features Selection | **1) Global similarity**<br>i) 2-gram encoding method<br>ii) 6-letter exchange group methods.<br><br>**2) Local similarity**<br>i) Len, Mut, and occur calculation.<br>ii) Min. description length (MDL) principle | i) Molecular weight (W)<br>ii) Isoelectric point (pI)<br>iii) Hydropathy composition (C)<br>iv) Hydropathy distribution (D)<br>v) Hydropathy transmission (T) | i) Sequence Arithmetic<br>ii) Reduce based Decision Tree (RDT)<br>iii) Neighbourhood Analysis (NA)<br>iv) Concept Lattice (CL) |
| Accuracy Level | 90% to 92% | 93% | 97.7% |
| Drawbacks | i) Better for Non-linear and Noisy data.<br>ii) Does not handle Physical relationship. | i) Concerned only about the physical Structure of AA.<br>ii) Does not handle Physical relationship. | i) No analytical output.<br>ii) Need Extra Time and Space |

## 4. PROPOSED MODEL

The aim of this model is to classify the unknown protein sequences in different families, classes or sub classes with high accuracy level and low computational time. We choose an unknown protein sequence as an input and extract some features from it and match with predefined values to classify the sequence into classes, sub classes and families. To do this first of all, extraction of features which is used to classify, is very important.

The proposed technique consists of three phases. The first phase aims to reduce the input dataset. The second phase helps to increase accuracy level of classification and the third phase implies the association rule to classify the protein sequence. Figure 1 gives a pictorial representation of the different modules of the proposed technique.

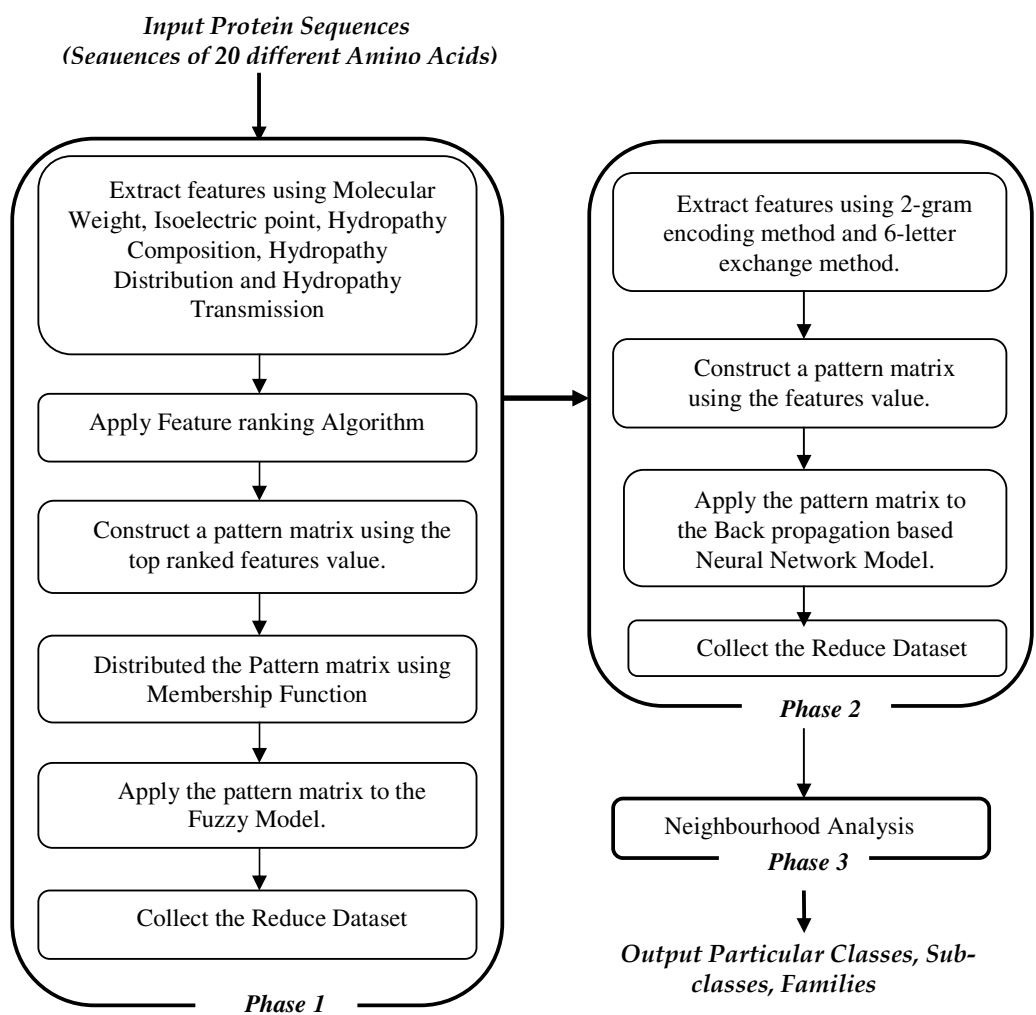

Figure 1. Pictorial representation of the different modules of the proposed technique

## 4.1. Phase 1

In first phase global features are extracted from the input protein sequence. These global features represent the nature of the entire protein sequence and global similarity between related sequences allowing for comparison. To extract the global features at first physico-chemical properties i.e. Molecular weight (W) and the isoelectric point (pI), of the input sequences are calculated. Apart of those features Hydropathy composition and Hydropathy distribution also calculated using the table 1.

Table 1: Chart of Molecular weight, isoelectric point, Hydropathy Property of 20 different Amino Acids

| Name | 1 letter Symbol | Molecular weight (W) | Isoelectric point (pI) | Hydropathy Property |
|---|---|---|---|---|
| Alanine | A | 89.10 | 6.00 | Hydrophobic |
| Arginine | R | 174.20 | 10.76 | Hydrophilic |
| Asparagine | N | 132.12 | 5.41 | Hydrophilic |
| Aspartic Acid | D | 133.11 | 2.77 | Hydrophilic |
| Cysteine | C | 121.16 | 5.07 | Hydrophobic |
| Glutamic Acid | E | 147.13 | 3.22 | Hydrophilic |
| Glutamine | Q | 146.15 | 5.65 | Neutral |
| Glycine | G | 75.07 | 5.97 | Neutral |
| Histicline | H | 155.16 | 7.59 | Neutral |
| Isoleucine | I | 131.18 | 6.02 | Hydrophobic |
| Leucine | L | 131.18 | 5.98 | Hydrophobic |
| Lysine | K | 146.19 | 9.74 | Hydrophilic |
| Methinine | M | 149.21 | 5.74 | Hydrophobic |
| Plenylalanine | F | 165.19 | 5.48 | Hydrophobic |
| Proline | P | 115.13 | 6.30 | Hydrophilic |
| Serine | S | 105.09 | 5.68 | Neutral |
| Threonine | T | 119.12 | 5.60 | Neutral |
| Tryptophan | W | 204.23 | 5.89 | Hydrophobic |
| Tyrosine | Y | 181.19 | 5.66 | Hydrophobic |
| Valine | V | 117.15 | 5.96 | Hydrophobic |

For an example Let consider a small protein sequence MARETFAR. According to the table 1 molecular weight of M is 149.21, A is 89.10, R is 174.20, E is 147.13, T is 119.12, F is 165.19, so total molecular weight of this sequence is (149.21 + 89.10 + 174.20 + 147.13 + 119.12 + 165.19 + 89.10 + 174.20) = 1107.25 and average molecular weight is 138.41. In the same way isoelectric point is calculated. Isoelectric point of M is 5.74, A is 6.00, R is 10.76, E is 3.22, T is 5.60, F is 5.48, so total isoelectric point of this sequence is (5.74 + 6.00 + 10.76 + 3.22 + 5.60 + 5.48 + 6.00 + 10.76) = 53.56 and average molecular weight is 6.69. Now consider the hydropathy property of the input sequence. According to the table M is hydrophobic in nature, A is hydrophobic, R is hydrophilic, E is hydrophilic, T is Neutral and F is Hydrophobic in nature. Table 2 and Table 3 show the Hydropathy composition and Hydropathy distribution of the input protein sequence.

Table 2. Result of Hydropathy Composition of Input Protein Sequence

| Hydropathy Composition | Occurrences | Percentage of Occurrences |
|---|---|---|
| Hydrophobic | 4 | 50.00% |
| Hydrophilic | 3 | 37.50% |
| Neutral | 1 | 12.50% |

Table 3. Result of Hydropathy Distribution of Input Protein Sequence

| Hydropathy Distribution | Occurrences | Percentage of Occurrences |
|---|---|---|
| Hydrophobic – Hydrophobic | 2 | 28.57 % |
| Hydrophobic – Hydrophilic | 2 | 28.57 % |
| Hydrophobic – Neutral | 0 | 00.00 % |
| Hydrophilic Hydrophobic | 0 | 00.00 % |
| Hydrophilic – Hydrophilic | 1 | 14.29 % |
| Hydrophilic – Neutral | 1 | 14.29 % |
| Neutral – Hydrophobic | 1 | 14.29 % |
| Neutral – Hydrophilic | 0 | 00.00 % |
| Neutral – Neutral | 0 | 00.00 % |

After extracting those features, a feature ranking algorithm will be applied on it. This algorithm will be able to provide a rank to the feature values and arrange the features values according to the descending order [8]. Top rank means have an extra ability to classify the protein sequence. After that pattern matrix will be generated. Now this pattern matrix will be distributed within three groups [6] and applied to the Fuzzy model. Those features which are used here generally deal with the molecular structure of the amino acids. So it is possible to eliminate huge no of families of protein in which the input protein sequence does not belongs. In this case data by data analysis was implemented instead of extracting knowledge based information. The main advantage of the data by data analysis is it provides the high accuracy level.

After completion of 1st Phase, generally the unknown input protein sequences may be classified in to particular families otherwise it definitely reduces the trained dataset. If phase 1 able to classified the input unknown protein sequence in to families, no need to continue this process further. The process will stop here otherwise the process will continue using 2nd Phase.

### 4.2. Phase 2

In 2nd Phase, 2-gram encoding method and 6-letter exchange group method both are used to extract the global similarity of the protein sequence [2]. These two methods are directly related to the structure of a protein sequence. In the second phase if back propagation technique of the neural network [4] is used on the reduce data set to extract the knowledge based information then more accrue result will be generated. Now this pattern matrix acts as the input of the neural network model. If dataset is huge then data by data analysis takes huge computational time. But here data set is small because phase 1 already reduced the data set. On the other hand extraction of knowledge based information provides more accurate result. So accuracy level is increase in this phase.

Here 2-gram encoding method was proposed which is described in [1]. The 2-gram encoding method extracts and counts the occurrences of patterns of two consecutive amino acids in a protein sequence. Let take the previous example of small protein sequence MARETFAR and apply 2-gram encoding method on it and provide the following result 1 for MA (occurrence of MA is 1 in the total sequence), 2 for AR (occurrence of AR is 2 in the total sequence), 1 for RE, 1 for ET, 1 for TA and 1 for FA. After that calculate the value for every pair using the formula x = c / (len(S) – 1), where x is the value, c is no of occurrence of every pair and len(S) is the total length of the input protein sequence, i.e. 0.1429 for MA, 0.2857 for AR and 0.1429 for others. Finally calculate the mean value (m) = 0.1667 and standard deviation (d) = 0.448 using the formula 1and 2

$$m = \frac{(\sum_{i=1}^{N} x_i)}{N} \quad \ldots\ldots\ldots\ldots\ldots\ldots\ldots\ldots\ldots\ldots (1)$$

$$d = \sqrt{\frac{(\sum_{i=1}^{N}(X_i - m)^2)}{(N-1)}} \quad \ldots\ldots\ldots\ldots\ldots\ldots\ldots\ldots\ldots\ldots\ldots\ldots (2)$$

In this phase 6-letter exchange group {e1,e2,e3,e4,e5,e6} also is adopted to represent a protein sequence, where e1 € {H,R,K}, e2 € {D,E,N,Q}, e3 € {C}, e4 € {S,T,P,A, G}, e5 € {M,I,L,V}, e6 € {F,Y,W}. For example, the above protein sequence MARETFAR can be represented as e5e4e1e2e4e6e4e1. The 6-letter exchange group method provides the following result 1 for e5e4 (occurrence of e5e4 is 1 in the total sequence), 2 for e4e1 (occurrence of e4e1 is 2 in the total sequence), 1 for e1e2, 1 for e2e4, 1 for e4e6 and 1 for e6e4. After that calculate the value for every pair using the formula  x = c / (len(S) – 1), where x is the value, c is no of occurrence of every pair and len(S) is the total length of the input protein sequence, i.e. 0.1429 for e5e4, 0.2857 for e4e1 and 0.1429 for others. Finally calculate the mean value (m) = 0.1667 and standard deviation (d) = 0.448 using the formula describe above.

For each protein sequence, both the 2-gram amino acid encoding and the 2-gram exchange group encoding are applied to the sequence. Thus, there are 20 X 20 + 6 X 6 = 436 possible 2-gram patterns in total. If all the 436 2-gram patterns are chosen as the neural network input features

After complete the 2$^{nd}$ Phase, generally the unknown input protein sequences may be classified in to particular families otherwise it definitely reduces the trained dataset. If phase 2 able to classified the input unknown protein sequence in to families, no need to continue this process further. The process will stop here otherwise the process will continue using 3$^{rd}$ Phase.

### 4.3.Phase 3

In the third phase, Neighbourhood Analysis (NA) will be used to classify the input sequence in the particular class or family. To use neighbourhood analysis we generally apply association rule. This rule has a power to extract the particular association between the protein sequence and classes, sub classes and families. So it is possible to eliminate all other classes, sub classes and families of protein in which this input protein sequence do not belongs

## 5. IMPLEMENTATION

The three different phases of the new proposed model have been implemented using an own designed tool and try to prove the incremental accuracy level and low computational time than others models. The proposed model is tested with various input protein sequences. In the next part description of tool and execution procedure are described.

### 5.1.Tool Description

A tool named Protein Sequence classifier is designed to implement the three different phases of proposed model. Here unknown protein sequence is placed as an input and classifies in to particular protein family with high accuracy level and low computational time. It is designed by JDK 1.6_16 and run on JAVA platform. A user friendly GUI is highly implemented here. Figure 2 & 3 show the phase 1 and phase 2 respectively. Another important part is also be incorporated with it i.e. this tool is able to insert the new data into data warehouse. Figure 4 shows the data insert phase.

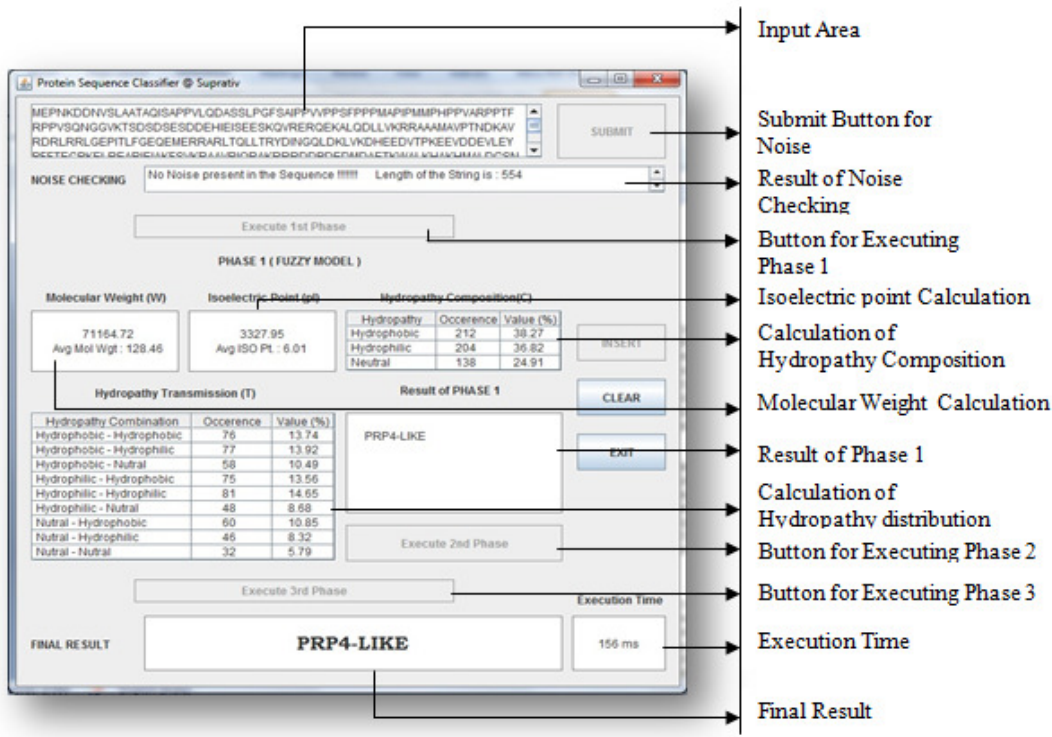

Figure 2. Screenshot of phase 1 execution in New designed tool.

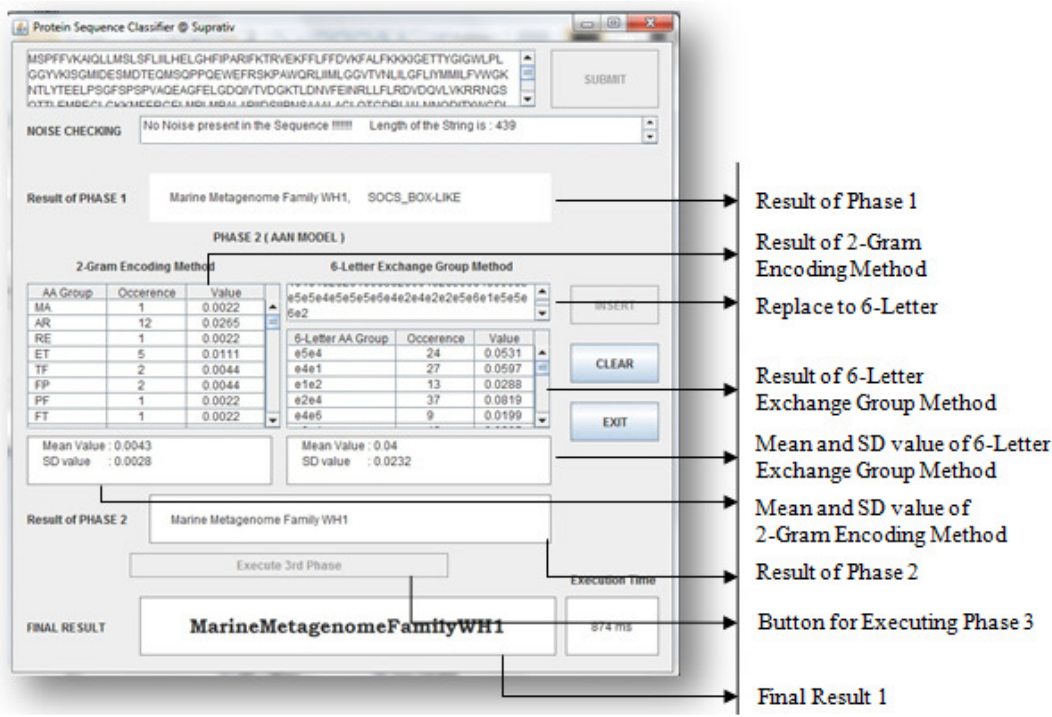

Figure 3. Screenshot of phase 2 execution in New designed tool.

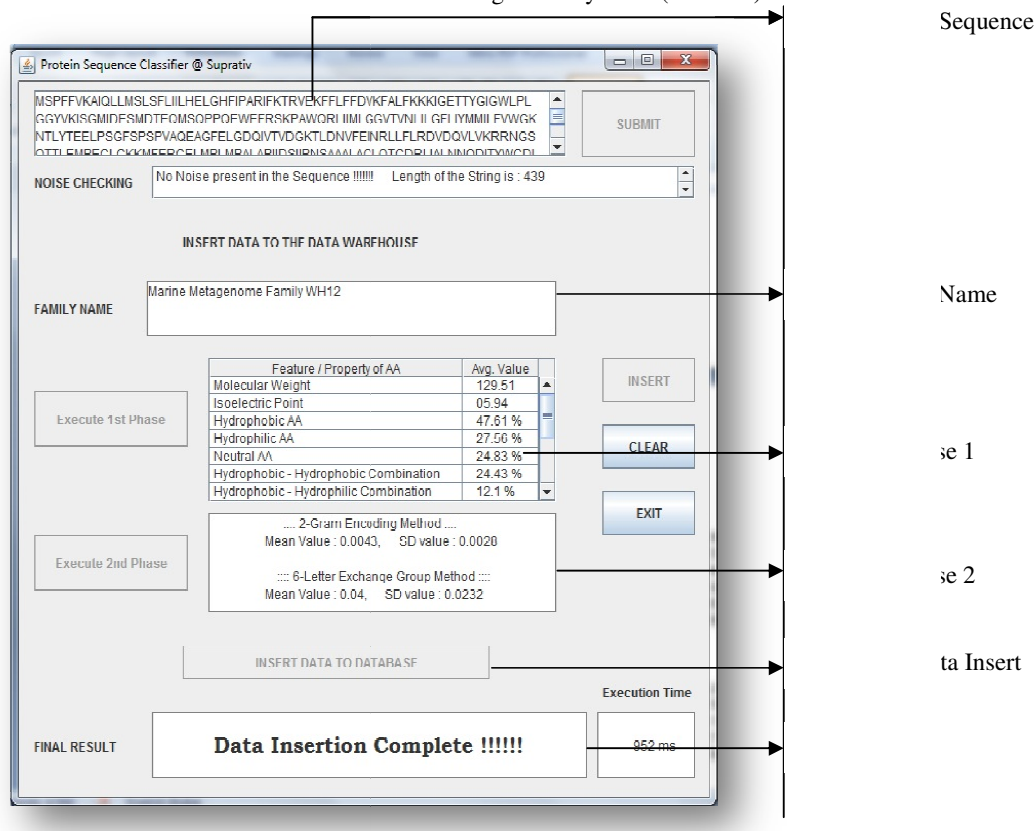

Sequence

Name

se 1

se 2

ta Insert

Figure 4. Screenshot of data Insertion module in New designed tool.

## 5.2. Execution Procedure

According to the previous description of the proposed model, total execution will be done in three different phases. At first user give an unknown protein sequence and submit for noise checking. If the input sequence is noise free then only two different doors will be opened to continue the further process i.e. 1) to execute phase 1 and 2) to insert data into the data warehouse, otherwise execution will be stopped. In the time of noise checking period, other some backend module also been executed. System creates database name knowledge with the help of data warehouse. Data warehouse stores the value of the every feature extracted in two different phases, against the protein family. Knowledge database store range of different feature values with respect to protein family. The overall process use this knowledge table instate of data warehouse to continue the further process.

Here another method also been incorporated to reduce the computational overhead. A single row column table named countrow is created to store the size of the data warehouse. Every time system count size of the data warehouse and match with the value of countrow, if value is different, then only knowledge table is created otherwise execute with the previous one.

Now go for execute Phase 1. Average Molecular Weight, Average Isoelectric point, Hydropathy composition and distribution are extracted and match with the data of Knowledge table. If system able to provide single family corresponding to the input, then process will be stop and family name will be placed in the final output box. Second phase will be executed if and only if the size of the phase 1 result set is greater than one. Otherwise execute phase 3 for prediction.

In the second phase 2-gram encoding method and 6 letter exchange group method are calculated and continue the matching procedure according to the previous one. In the third phase neighborhood analysis will be done for prediction. A detail flow diagram is given in figure 5.

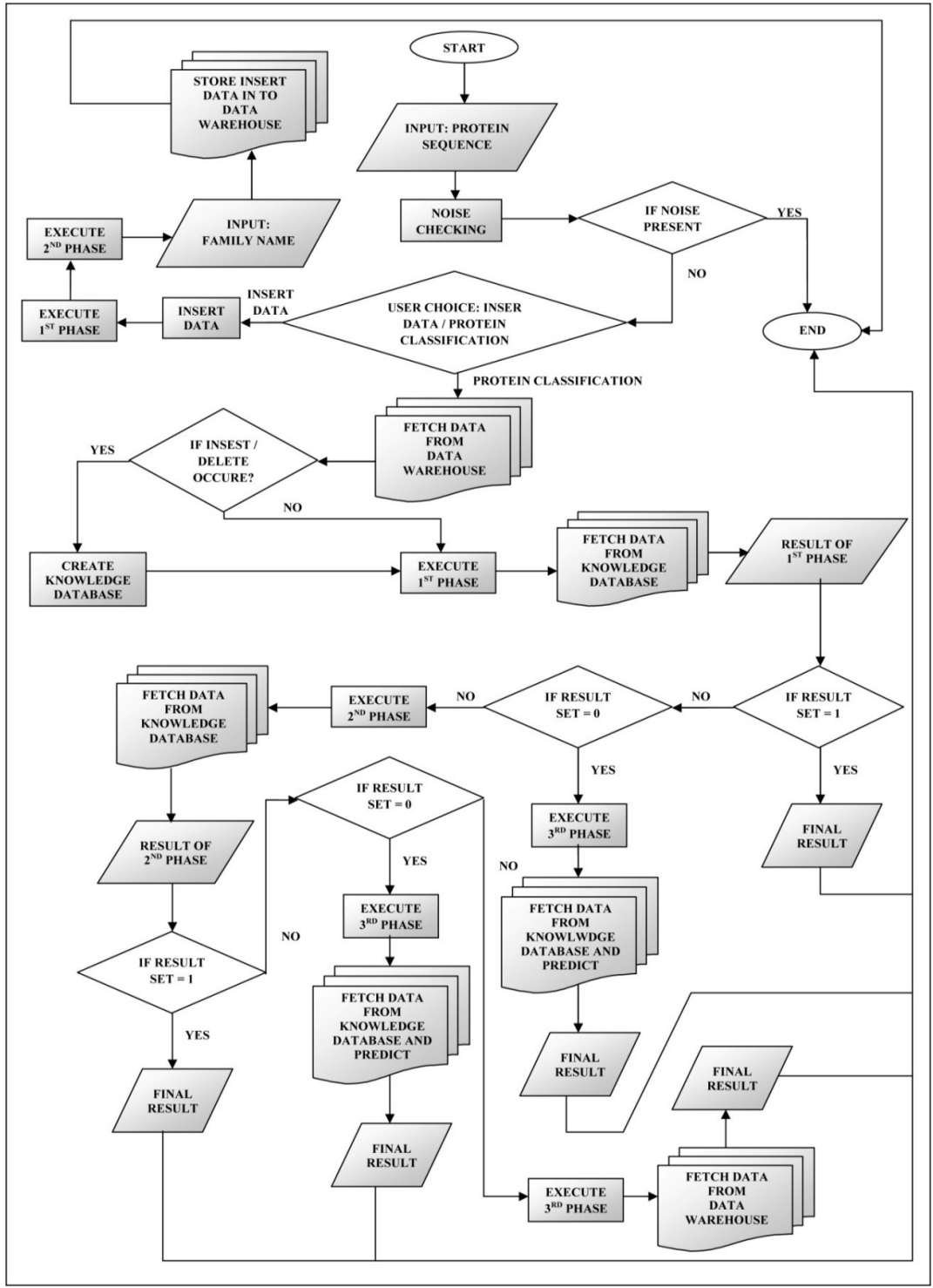

Figure 5. A Flow Diagram to execute the proposed classifier

Another module also be attached which can able to insert data in to data warehouse. After completing the noise checking procedure execute phase 1 and phase 2 one after another and provide the family name. Finally click on insert data button, system store all the information in to the data warehouse.  In the next a flow diagram is designed to describe the overall process.

## 6. RESULT DESCRIPTION

A series of experiments are carried out to evaluate the performance of the proposed new classifier technique named "*Protein Sequence Classifier*" on a Pentium IV PC running the Windows XP operating system. Generally the data used to construct the data warehouse for testing the classifier tool were obtained from NCBI and SCOP. Data means different protein sequences were obtained from NCBI and corresponding family name of these protein sequences were got from SCOP. After that this classifier tool helps us to create the data warehouse using the inserting feature of this classifier tool.

As a testing dataset five positive datasets of protein families are considered there are 1) FaeA-LIKE, 2) Marine Metagenome Family WH1, 3) MiaE-LIKE, 4) PRP4-LIKE and 5) SOCS_BOX-LIKE. At least 500 to 800 sequences of different length of every family are taken to calculate the features which were stored in the data warehouse. Means feature values of total 2500 to 4000 sequences are used to construct the data warehouse as a testing purpose. As a result 2500 to 4000 rows are available and total 18 features are considered for every protein sequences. So total size of the data warehouse is 19 * 4000. [18 features and 1 column to store the corresponding family name so 19 columns and 4000 rows]

In this classifier with in three phases one phase execute Fuzzy based model, second one execute Neural Network based Model and another one is neighborhood Analysis. After evaluating the performance a huge success is got to execute the Fuzzy based model before execution of Neural based model. Generally it also proved that after analyzing 500 unknown protein sequences 395 protein sequences are classified using only Fuzzy based model. In these 395 cases we do not need to execute the Neural based model so timing complexity is reduced. Another 75 cases need to execute the Neural Network based model after executing the Fuzzy based Model. And left 30 cases need to execute all three cases. So from this analysis that is concluded, Fuzzy based Model are considered as the 1$^{st}$ Phase, Neural Network based model are execute in next and Finally Neighborhood analysis are performed.

Let take an example of performance evaluation program. An unknown protein sequence is taken as an input and executes it in two different classifiers.

*The input Protein Sequence:*

```
MEPNKDDNVSLAATAQISAPPVLQDASSLPGFSAIPPVVPPSFPPPMAPIPMMPHPPVARPPTFRPPVSQ
NGGVKTSDSDSESDDEHIEISEESKQVRERQEKALQDLLVKRRAAAMAVPTNDKAVRDRLRRLGEPITLF
GEQEMERRARLTQLLTRYDINGQLDKLVKDHEEDVTPKEEVDDEVLEYPFFTEGPKELREARIEIAKFSV
KRAAVRIQRAKRRRDDPDEDMDAETKWALKHAKHMALDCSNFGDDRPLTGCSFSRDGKILATCSLSGVTK
LWEMPQVTNTIAVLKDHKERATDVVFSPVDDCLATASADRTAKLWKTDGTLLQTFEGHLDRLARVAFHPS
GKYLGTTSYDKTWRLWDINTGAELLLQEGHSRSVYGIAFQQDGALAASCGLDSLARVWDLRTGRSILVFQ
GHIKPVFSVNFSPNGYHLASGGEDNQCRIWDLRMRKSLYIIPAHANLVSQVKYEPQEGYFLATASYDMKV
NIWSGRDFSLVKSLAGHESKVASLDITADSSCIATVSHDRTIKLWTSSGNDDEDEEKETMDIDL
```

In the first one (fig: 6) Fuzzy based Model are considered as the 1$^{st}$ Phase, Neural Network based model are execute in next and Finally Neighborhood analysis are performed and the second one (fig: 7) Neural Network based model are considered as the 1$^{st}$ Phase, Fuzzy based Model are execute in next and Finally Neighborhood analysis are performed.

*Output:* PRP4-like

*Execution Time of 1st Classifier:* 156 ms (Show in Figure 6)

*Execution Time of 2nd Classifier:* 1296 ms (Show in Figure 7)

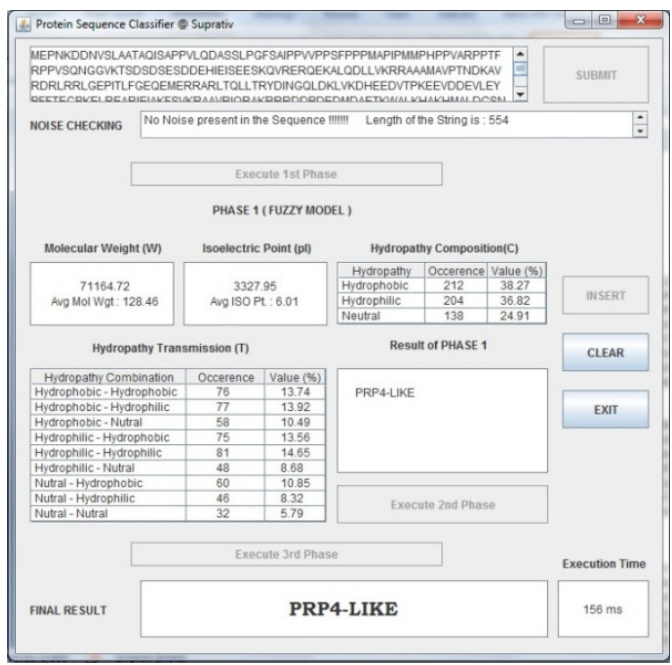

Figure 6. Screenshot where Fuzzy based model is executing first

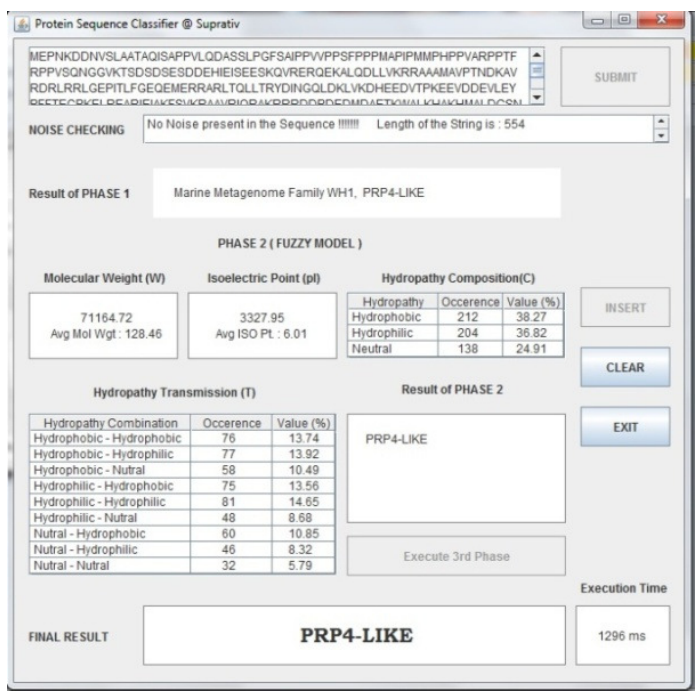

Figure 7. Screenshot where Neural Based Model is executing first.

In figure 6 and figure 7 tool execute on same input data and access same database but in figure 6, the execution time is less than figure7 because in first one tool takes only one phase to produce output but in the second figure tool takes 2 phases to compute require result. So it can be commented that if fuzzy model is implemented as the first phase then tools take less time to compute and it may chance it can compute using only single phase.

## 7. CONCLUSION

Protein analysis has a major part in the domain of biological research. So before analyzing, identification of protein is mostly needed. As a result different protein sequence classification techniques are invented, this has enhance the research scope in Biological Domain. Choosing the perfect techniques regarding this classification is also a most important part of the research.

Recent trend analysis shows that it is very difficult to classify large amount of biological data like protein sequences using traditional database system. Data mining technique are appropriate to handle the large data sources. A number of different techniques are prevalent for classifying protein sequences. This dissertation includes a detail review of ongoing research work involving three different techniques to classify the protein sequences. A comparative study is done to understand the basic difference between these models. The accuracy level of each proposed model has been studied. Those brief review and comparative study help to understand the insufficiency of the previous classification techniques with respect to the accuracy level and computational time.

A new classification model has been proposed in this dissertation for classifying the unknown protein sequences into families producing knowledge based information beside data analysis technique. The classifier has two phases, (i) Fuzzy Logic based Classifying (ii) Neural Network cased Classifying. In this new proposed model Fuzzy based algorithm and Neural network based algorithm both are incorporated. Fuzzy logic based algorithm helps us to do the data by data analysis which is very useful to reduce the total dataset and neural network based algorithm produce a knowledge by analysis the overall data warehouse which help us to classify and predict the require output. Neighborhood analysis is done in the final stage of the new proposed technique for predicting the required output.

The proposed classifier has been implemented using JAVA (JDK 1.6_16). We have also tested the model with the order of phases reversed. The model has been tested using a test data of 500 different sequences. It has been observed that the accuracy level increase if the fuzzy based logic is executed before the neural network based logic.

## REFERENCES

[1] Cathy Wu, Michael Berry, Sailaja Shivakumar , Jerry Mclarty (1995) Neural Networks for Full-Scale Protein Sequence Classification: Sequence Encoding with Singular Value Decomposition. Kluwer Academic Publishers, Boston. Manufactured in The Netherlands, Machine Learning, 21, pp: 177-193.

[2] Jason T. L. Wang, Qic Heng Ma, Dennis Shasha, Cathy H Wu (2000) Application of Neural Networks to Biological Data Mining: A case study in Protein Sequence Classification. KDD, Boston, MA, USA, pp: 305-309.

[3] C. Z. Cai, L. Y. Han, Z. L. Ji, X. Chen, and Y. Z. Chen (2003) SVM-Prot: Web-based support vector machine software for functional classification of a protein from its primary sequence. Nucleic Acids Research, vol. 31, pp. 3692–3697.

[4] Dianhui Wang, Guang-Bin Huang (2005) Protein Sequence Classification Using Extreme Learning Machine. Proceedings of International Joint Conference on Neural Networks (IJCNN2005), Montreal, Canada.

[5] Shakir Mohamed, David Rubin and Tshilidzi Marwala (2006) Multi-class Protein Sequence Classification Using Fuzzy ARTMAP. IEEE Conference pp: 1676 – 1680.

[6] E. G. Mansoori, M. J. Zolghadri, S. D. Katebi, H. Mohabatkar, R. Boostani and M. H. Sadreddini (2008) Generating Fuzzy Rules For Protein Classification. Iranian Journal of Fuzzy Systems Vol. 5, No. 2, pp. 21-33.

[7] Zarita Zainuddin, Maragathan Kumar (2008) Radial Basic Function Neural Networks in Protein Sequence Classification. Malaysian Journal of Mathematical Science2(2), pp: 195-204.

[8] Eghbal G. Mansoori, Mansoor J. Zolghadri, and Seraj D. Katebi (March 2009) Protein Superfamily Classification Using Fuzzy Rule-Based Classifier. IEEE Transactions On Nanobioscience, Vol. 8, No. 1, pp 92-99.

[9] PV Nageswara Rao, T Uma Devi, Dsvgk Kaladhar, Gr Sridhar, Allam Appa Rao (2009) A Probabilistic Neural Network Approach For Protein Superfamily Classification. Journal of Theoretical and Applied Information Technology.

[10] Ramadevi Yellasiri, C.R.Rao (2009) Rough Set Protein Classifier. Journal of Theoretical and Applied Information Technology.

[11] Shuzlina Abdul Rahman, Azuraliza Abu Bakar, Zeti Azura Mohamed Hussein (2009) Feature Selection and Classification of Protein Subfamilies Using Rough Sets. International Conference on Electrical Engineering and Informatics, Selangor, Malaysia.

[12] Suprativ Saha, Rituparna Chaki (2012) "A Brief Review of Data Mining Application Involving Protein Sequence Classification", Advances in Intelligent Systems and Computing, 1, Volume 177, Advances in Computing and Information Technology, Springer Publisher, ACITY 2012, Chennai, India, pp. 469-477

[13] George Tzanis, Christos Berberidis, and Ioannis Vlahavas, "Biological Data Mining".

**Authors**

Suprativ Saha is Pursuing M.E. in Computer Science and Engineering from West Bengal University of Technology. He is presently working as an Assistant Professor at Global Institute of Management and Technology, Krishnagar City, West Bengal, India. His research interests include the field of Database Management System, Data Mining and Distributed Database. His area of research is Biological Data Mining.

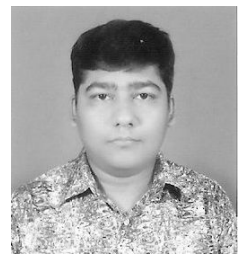

Rituparna Chaki is an Associate Professor in the Department of Computer Science & Engineering, West Bengal University of Technology, Kolkata, India, since 2005. She received her Ph.D. in 2002 from Jadavpur University, India. The primary area of research interest for Dr. Chaki is Wireless Mobile Ad hoc Networks and Data Mining. She has also served as a Systems Manager for Joint Plant Committee, Government of India for several years before she switched to Academia. Dr. Chaki also serves as a visiting faculty member in other leading Universities including Jadavpur University. Dr. Chaki has about 45 referred international publications to her credit.

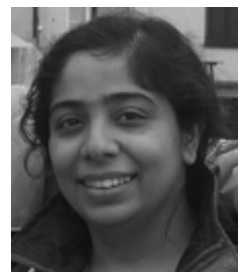